\documentstyle[amssymb,11pt,multicol,epsf]{article} 

\begin{document} 



\title{\vspace{-40pt}$\phantom{}$\\ The shortest path to complex networks}

\author{S. N. Dorogovtsev\,$^{\ast,\dag,\ddag}$ and J. F. F. \vspace{4pt}Mendes\,$^{\ast}$ 
\\
{\small \it Physics Department of the University of \vspace{-1pt}Aveiro}, 
\\ 
{\small \it Laboratory of Physics at the Helsinki University of \vspace{-1pt}Technology}, 
\\
{\small \it A.F. Ioffe Physico-Technical Institute, \vspace{-16pt}St Petersburg}
}

\date{}

\maketitle

\begin{abstract}{}
In these introductory notes for `pedestrians' we describe the current state of the art in the science of complex networks. 
\end{abstract} 


\bibliographystyle{plain}



\centerline{\small \bf \vspace{-5pt}Contents}

\begin{multicols}{2}

\begin{footnotesize}

\begin{list}{}{\leftmargin=18pt}

\item[$\phantom{.}$\,1.\,\ ] 
The birth of network \vspace{-3pt}science 
\item[$\phantom{.}$\,2.\,\ ] 
What are random \vspace{-3pt}networks?  
\item[$\phantom{.}$\,3.\,\ ]
Adjacency \vspace{-3pt}matrix 
\item[$\phantom{.}$\,4.\,\ ]
Degree \vspace{-3pt}distribution 
\item[$\phantom{.}$\,5.\,\ ]
What are simple \vspace{-2pt}networks? Classical random \vspace{-3pt}graphs 
\item[$\phantom{.}$\,6.\,\ ]
The birth of the giant \vspace{-3pt}component 
\item[$\phantom{.}$\,7.\,\ ]
Topology of the \vspace{-3pt}Web 
\item[$\phantom{.}$\,8.\,\ ]
Uncorrelated \vspace{-3pt}networks 
\item[$\phantom{.}$\,9.\,\ ]
What are small \vspace{-3pt}worlds? 
\item[$\phantom{.}$\,10.\,\ ]
Real networks are \vspace{0pt}mesoscopic \vspace{-3pt}ob\-jects 
\item[$\phantom{.}$\,11.\,\ ]
What are complex \vspace{-3pt}networks? 
\item[$\phantom{.}$\,12.\,\ ]
The configuration \vspace{-3pt}model 
\item[$\phantom{.}$\,13.\,\ ]
The absence of degree--degree correlati\vspace{-3pt}ons 
\item[$\phantom{.}$\,14.\,\ ]
Networks with correlated \vspace{-3pt}degrees 
\item[$\phantom{.}$\,15.\,\ ]
Cluster\vspace{-3pt}ing
\item[$\phantom{.}$\,16.\,\ ]
What are small-world \vspace{-3pt}networks? 
\item[$\phantom{.}$\,17.\,\ ]
`Small worlds' is not \vspace{-2pt}the same as `small-world \vspace{-3pt}networks' 
\item[$\phantom{.}$\,18.\,\ ]
Fat-tailed degree \vspace{-3pt}distributions 
\item[$\phantom{.}$\,19.\,\ ]
Reasons for the fat-tailed degree distri\vspace{-3pt}butions  
\item[$\phantom{.}$\,20.\,\ ] 
Preferential \vspace{-3pt}linking 
\item[$\phantom{.}$\,21.\,\ ]
Condensation of \vspace{-3pt}edges 
\item[$\phantom{.}$\,22.\,\ ]
Cut-offs of degree \vspace{-3pt}distributions 
\item[$\phantom{.}$\,23.\,\ ]
Reasons for correlations in \vspace{-3pt}networks 
\item[$\phantom{.}$\,24.\,\ ]
Classical random graphs \vspace{-2pt}cannot be used for comparison with real net\-wo\vspace{-3pt}rks 
\item[$\phantom{.}$\,25.\,\ ]
How to measure \vspace{-2pt}degree--degree correlati\vspace{-3pt}ons  
\item[$\phantom{.}$\,26.\,\ ]
Assortative and disassortative mix\-i\vspace{-3pt}ng 
\item[$\phantom{.}$\,27.\,\ ]
Disassortative mixing does \vspace{-2pt}not mean that vertices of high degrees 
\vspace{-2pt}rarely connect to each \vspace{-3pt}other 
\item[$\phantom{.}$\,28.\,\ ]
Reciprocal links in directed net\-wo\vspace{-3pt}rks 
\item[$\phantom{.}$\,29.\,\ ]
Ultra-small-world \vspace{-3pt}effect 
\item[$\phantom{.}$\,30.\,\ ]
The importance of the tree ansa\vspace{-3pt}tz 
\item[$\phantom{.}$\,31.\,\ ]
Ultraresilience against random fai\-lu\-\vspace{-3pt}res 
\item[$\phantom{.}$\,32.\,\ ]
When correlated nets are ultra\-re\-si\-li\vspace{-3pt}ent 
\item[$\phantom{.}$\,33.\,\ ]
Vulnerability of complex \vspace{-3pt}networks 
\item[$\phantom{.}$\,34.\,\ ]
The absence of an epidemic thre\-s\-\vspace{-3pt}hold 
\item[$\phantom{.}$\,35.\,\ ]
Search based on local informati\vspace{-3pt}on 
\item[$\phantom{.}$\,36.\,\ ]
Ultraresilience disappears \vspace{-2pt}in finite network\vspace{-3pt}s 
\item[$\phantom{.}$\,37.\,\ ]
Critical behavior of \vspace{-2pt}cooperative mo\-dels on networ\vspace{-3pt}ks 
\item[$\phantom{.}$\,38.\,\ ]
Berezinskii-Kosterlitz-\vspace{-2pt}Thouless phase transitions in \vspace{-3pt}networks 
\item[$\phantom{.}$\,39.\,\ ]
Cascading \vspace{-3pt}failures  
\item[$\phantom{.}$\,40.\,\ ]
Cliques \vspace{-2pt}and communiti\vspace{-2pt}es 
\item[$\phantom{.}$\,41.\,\ ]
Betweenn\vspace{-3pt}ess 
\item[$\phantom{.}$\,42.\,\ ]
Extracting \vspace{-3pt}communities 
\item[$\phantom{.}$\,43.\,\ ]
Optimal \vspace{-3pt}paths 
\item[$\phantom{.}$\,44.\,\ ]
Distributions of the \vspace{-2pt}shortest-path len\-gth and of the loop's length are 
\vspace{-3pt}narrow 
\item[$\phantom{.}$\,45.\,\ ]
Diffusion on \vspace{-4pt}networks 
\item[$\phantom{.}$\,46.\,\ ]
What is \vspace{-3pt}modularity? 
\item[$\phantom{.}$\,47.\,\ ]
Hierarchical organization of net\-wo\vspace{-3pt}rks 
\item[$\phantom{.}$\,48.\,\ ]
Convincing modelling of \vspace{-2pt}real-world networks: Is it \vspace{-3pt}possible? 
\item[$\phantom{.}$\,49.\,\ ]
The small \vspace{-3pt}Web  
\item[$\phantom{.}$\,50.\,\ ]
The failures and perspectives \vspace{-2pt}of the physics approach to complex \vspace{-3pt}networks 
\item[$\phantom{.}$\,51.\,\ ]
A remark about references 

\end{list}

\end{footnotesize}

\end{multicols}


\renewcommand{\thesubsection}{\arabic{subsection}} 


\subsection{The birth of network science}\label{ss-birth_network_science}

In 1735, in St Petersburg, Leonhard Euler solved the so-called K\"onigsberg bridge problem---walks on a simple small graph. This solution (actually, a proof) is usually considered as a starting point of the science of networks.\footnote{
For recent reviews and reference books, see Refs.~[1--9].  
}

\subsection{What are random networks?}\label{ss-random_networks}

Random networks are {\em statistical ensembles of graphs}. 
A statistical ensemble is a set of its members---particular graphs, each of which has its specific probability of realization---a statistical weight. 

In empirical studies, as a rule, a single member (a particular realization) of this ensemble is observed. In simulations, a finite number of realizations of the ensemble may be obtained. 


As is standard in statistical mechanics, statistical ensembles are classified as equilibrium or non-equilibrium. In our case, these are equilibrium and non-equilibrium (e.g., growing) random networks.

\subsection{Adjacency matrix}\label{ss-adjacency_matrix}

The complete description of a particular graph is provided by its adjacency matrix. A graph of $N$ vertices has an $N\times N$ adjacency matrix. Each element $a_{ij}$ of the adjacency matrix is equal to the number of edges connecting the vertices $i$ and $j$.

\subsection{Degree distribution}\label{ss-degree}

The simplest local characteristic of a vertex is its {\em degree}: the total number of the edges attached to a vertex, that is the total number of the nearest neighbours of a vertex. 

In {\em directed networks}, the number of incoming/outgoing edges of a vertex is called its in-/out-degree. 

In a random network, a degree distribution is the average fraction of vertices of degree $k$: $P(k) = \langle N(k) \rangle /N$. Here $N(k)$ is the number of vertices of degree $k$ in a particular graph of the statistical ensemble. The averaging is over the entire statistical ensemble. 

An empirical researcher measures $N(k)/N$ for a single realization of the statistical ensemble. Simulations usually allow to average $N(k)/N$ over a finite set of realizations of the statistical ensemble.

\subsection{What are simple networks? Classical random graphs}\label{ss-simple_networks}

The simplest random networks are so-called classical random graphs 
(Solo\-mo\-noff and Rapoport, 1950--1952, Erd\H os and R\'enyi, 1959,1960, Gilbert 
1959). In simple terms, these are maximally random networks under the constraint that the mean degree of their vertices, $\langle k \rangle$, is fixed. (We assume that the number of vertices in these graphs is also fixed.) The maximum randomness means the maximum entropy of a random net. 

There are two main versions of classical random graphs:  

{\em The Erd\H os and R\'enyi model} (a widespread term) is a statistical ensemble of all possible graphs of precisely $N$ vertices and precisely $L$ edges, where each member of the ensemble has equal probability of realization. 

In {\em the Gilbert model}, each pair of $N$ vertices is connected with some probability $p$. This produces a statistical ensemble of all possible graphs of $N$ vertices. The members of these ensemble are weighted with some statistical weights. 
In the thermodynamic limit (infinitely large networks), these two versions are equivalent ($\langle k \rangle = p(N-1)$). 

The degree distribution of classical random graphs has a Poisson form: 
$P(k) \sim \langle k \rangle^k/k!$. Here $\langle k \rangle$ is fixed as $N \to \infty$. This is an extremely rapidly decreasing distribution with a natural scale $k \sim \langle k \rangle$. All its moments converge.

\subsection{The birth of the giant component}\label{ss-birth_of_giant}

The limit with a fixed $\langle k \rangle$ as $N \to \infty$ corresponds to {\em a sparse network}. In a sparse net, the mean number of connections of a vertex is much less than the number of connections of a vertex in a fully connected graph. 

Why is the case of a sparse network most interesting? 
The important feature of a network is its {\em giant connected component}. This is a set of mutually reachable vertices containing a finite fraction of vertices of a large network. 
Without the giant connected component, a net is only a set of small separated clusters.   
It turns out that 
in the classical random graphs, the giant connected component exists if the mean number of connections of a vertex exceeds one. So, this characteristic point of a networks---the point of `the birth of the giant connected component'---is just in the range of extremely low degrees, $\langle k \rangle \sim 1 \ll N$.

\subsection{Topology of 
the Web}\label{ss-topology_directed}

\begin{figure}
\epsfxsize=55mm
\centerline{\epsffile{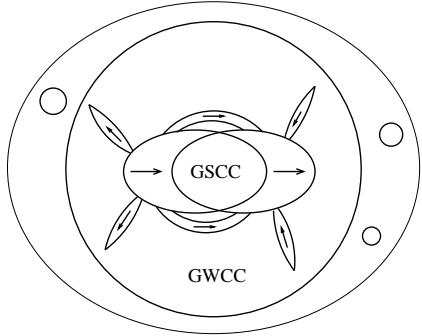}}
\caption{
\small{
The global structure of a directed network. 
The giant weakly connected component (GWCC), which is the giant connected component of the undirected projection of this network, contains: (1) giant strongly connected component (GSCC), (2) giant in-component, (3) giant out-component, and (4) `tendrils'. 
}  
}
\label{f1}
\end{figure}

{\em Directed networks} are networks with directed edges. 
As Fig.~\ref{f1} shows, these networks, generally, have a far more rich global structure than undirected ones. There are several types of giant connected components in directed nets. The core of a directed network is its {\em giant strongly connected component}, which consists of vertices mutually reachable by directed paths. 


Note that the scheme in Fig.~\ref{f1} is applicable, in general, to any directed network, including lattices. In particular, the WWW has this global structure \cite{bkm00}.

\subsection{Uncorrelated networks}\label{ss-uncorrelated_networks}

In principle, connected vertices (i.e., vertices, for which a connecting path exists) 
may be correlated. The examples of these correlations are (1) loops and (2) correlations between degrees of connected vertices (e.g., the nearest neighbours). 

Evidently, in large classical random graphs, correlations are absent: degrees of connected vertices are uncorrelated, and loops are not essential in the large network limit. Such (equilibrium) random networks are called uncorrelated. 

The fact that loops are not essential in the thermodynamic limit implies that any finite neighbourhood of a vertex has {\em a tree-like structure}.


\subsection{What are small worlds?}\label{ss-small_worlds} 

Accounting for the last circumstance allows us easily estimate the `linear size' of a classical random graph, that is {\em the mean length of the shortest path between two vertices}. Evidently, the mean number of the $n$-th nearest neighbours of a vertex rapidly grows as $\langle k \rangle^n$. So, the average shortest-path length $\overline{\ell}$ is roughly estimated by the condition $\langle k \rangle^{\overline{\ell}} \sim N$. Then, 

\begin{equation}
\overline{\ell} \approx \frac{\ln N}{\ln \langle k \rangle}
\, .  
\label{1}
\end{equation} 
This formula is asymptotically exact in classical random graphs. It also works well in random networks, where degree distributions rapidly decrease with $k$. 

Compare formula (\ref{1}) with the expression 
$\overline{\ell} \sim N^{1/d}$ for a linear size of a $d$-dimensional lattice. 
In networks, the size dependence $\overline{\ell}(N)$ is slower than in any finite-dimensional lattice or fractal. 

The growth of $\overline{\ell}(N)$ slower than any positive power of $N$ is called {\em a small-world effect}. By definition, a network is {\em a small world} if it shows the small-world effect. One can see that {\em small-worlds are infinite-dimension objects}. 
This feature is a basic property of networks.

\subsection{Real networks are mesoscopic objects}\label{ss-mesoscopic_objects}

Formally speaking, it is hard to make a solid conclusion that a given real-world network displays a small-world effect. 
The reason is that, as a rule, real networks are small (small numbers of vertices): 
there are $10^2$--$10^3$ vertices in most of empirically studied biological networks, about $2\times 10^4$ vertices in the Internet at the Autonomous Systems level, several hundreds thousands routers in the Internet, and `only' about $10^{10}$ pages in the WWW. 

These numbers are not large enough to check the small-world effect. 
Moreover, these numbers are not large enough to
treat networks as macroscopic systems, where the measurement of a small fraction of a system allows to arrive at complete conclusions about the entire system. 

Real networks are mesoscopic objects.\footnote{
A `microscopic' network is an edge connecting two vertices. 
} 
That is, a whole system (or, at least, its essential part) must be explored to arrive at the `complete' knowledge of the system. 
Note that, as a rule, the information about a whole real-world network (a full map, a complete adjacency matrix) is available. Usually, the complete map of a network is a starting point of the analysis of empirical data.

\subsection{What are complex networks?}\label{ss-complex_networks}

Complex random networks are networks which are more complex than classical random graphs. Here the term `complex' means more complex organization (more complex distribution of connection). In particular, a degree distribution may be more complex than Poisson, and/or various correlations may be essential.  

Real-world nets are complex networks, usually with fat-tailed degree distributions, usually with strong correlations of degrees of connected vertices, usually with an essential
role of loops.

\subsection{The configuration model}\label{ss-configuration_model}

The configuration model (the term is introduced by B.~Bollob\'as) is the first natural generalization of classical random graphs. In very simple terms, the configuration model is a maximally random graph with a given degree distribution $P(k)$. 

This complex random equilibrium network (recall---an ensemble!) is uncorrelated. 
Most of results for complex networks are obtained by using the configuration model.\footnote{
There is another, more traditional for statistical mechanics, way to build ensembles of networks. Sometimes, it is referred to as {\em the exponential model}. 
The members of the statistical ensemble in this construction are systems (sets) of local configurations of vertices and edges. Each kind of these clusters (`bricks')   
has its `excitation energy'. By thermal excitation one can obtain a full set of realizations (networks) of the ensemble. The specific excitation energies determine the statistical weights of these realizations, that is the structure of the resulting random network.   
}

\subsection{The absence of degree--degree 
correlations}\label{ss-absence_correlations}

What does this mean? 
In particular, this means that the degrees of the nearest-neighbour vertices uncorrelated. That is, the joint probability $P(k,k')$ that an edge connects vertices of degrees $k$ and $k'$ is 

\begin{equation}
P(k,k') = \frac{kP(k)k'P(k')}{\langle k \rangle^2}
\, .  
\label{2}
\end{equation} 
Indeed, each end of an edge with equal probability may occur to be in any of 
$2L = \langle k \rangle\sum_k k P(k)$ possible positions. So, an end of an edge in an uncorrelated network is attached to a vertex of degree $k$ with probability $kP(k)/\langle k \rangle$.

\subsection{Networks with correlated degrees}\label{ss-correlated_degrees}

The simplest case of degree--degree correlations are correlations between the degrees of the nearest-neighbour vertices. If these correlations are present in a network, the joint distribution $P(k,k')$ differs from the right-hand side of relation (\ref{2}). 

The configuration model may be generalized to include these correlations. 
The resulting correlated network is a maximally random graph under the restriction that the joint degree--degree distribution $P(k,k')$ is equal to a given function. (The degree distribution follows from $P(k,k')$ and so is also fixed.) 

The only type of correlations in the resulting network are correlations between the degrees of the nearest-neighbour vertices. So, the network also has a tree-like local structure. 

Proceeding in this way one can construct networks with more and more complex correlations between degrees of connected vertices.

\subsection{Clustering}\label{ss-clustering}

Loops are specific correlations in networks. The notion of clustering is related to loops of length three (triangles of edges). {\em The local clustering} is the relative number of connections between the nearest neighbours of a vertex $i$ 

\begin{equation}
C_i = \frac{n_i}{k_i(k_i-1)/2}
\, .  
\label{3}
\end{equation}  
Here $k_i$ is the degree of the vertex, $n_i$ is the total number of connections between its nearest neighbours. Averaging $C_i$ over vertices of degree $k$ provides {\em the degree-dependent local clustering} $C(k)$, which shows the probability that two nearest neighbours of a vertex of degree $k$ are connected to each other. 

{\em The mean clustering} is defined as 

\begin{equation}
\langle C \rangle \equiv \langle C_i \rangle = \sum_k P(k) C(k)
\, .  
\label{4}
\end{equation}  
Another clustering characteristic, {\em the clustering coefficient}, is defined as 

\begin{equation}
C \equiv \frac{\langle n_i \rangle}{\langle k_i(k_i-1)/2 \rangle} = 
\frac{\sum_k P(k)\langle n(k) \rangle}{(\langle k^2 \rangle - \langle k \rangle)/2}
\, .  
\label{5}
\end{equation}
The clustering coefficient is three times the ratio of the total number of edge triangles and the total number of the connected triples of vertices. 
Note that if the local clustering is degree-dependent, $\langle C \rangle \neq C$. The relative difference may be great in real-world networks.  

As is natural, in infinitely large uncorrelated networks, clustering is absent. 
So, in uncorrelated networks, the clustering is only a finite size effect. 
For example, in the classical random graphs, 

\begin{equation}
C(k) = C = \langle C \rangle \cong \frac{\langle k \rangle}{N}
\, .  
\label{6}
\end{equation} 
In the configuration model, 

\begin{equation}
C(k) = C =  \langle C \rangle \cong \frac{(\langle k^2 \rangle - \langle k \rangle)^2}{N\langle k \rangle^3}
\, .  
\label{7}
\end{equation} 

In networks without degree--degree correlations, the local clustering is degree independent, and $C =  \langle C \rangle$, but this is a rare exception. Formulae for $C(k)$, $C$, and $\langle C \rangle$ in networks with degree--degree correlations are given in Ref.~\cite{d04}. 

We believe that empirical data on clustering is usually determined by the form of $P(k)$ and $P(k,k')$. So, the strong enough clustering may be explained by using formula (\ref{7}) and corresponding expressions for networks with degree--degree correlations, 
without implementing some specific mechanism of strong clustering.

\subsection{What are small-world networks?}\label{ss-small_world_networks}

Nevertheless, one has to admit that   
as a rule, real-world networks have really strong clustering. Moreover, the values of the clustering coefficient are so high in some networks, that it is hard to believe that it is a finite-size effect. Watts and Strogatz proposed a specific class of complex networks, which have a small-world effect. 
They have named them small-world networks. These are lattices with high clustering (e.g., a trigonal lattice), where randomly chosen vertices are connected by long-range shortcuts. 

Actually, a small-world network is a superposition of a lattice and a classical random graph. Due to the strong clustering of a mother lattice, a small-world network has high clustering. Due to the compactness of the classical random graph, a small-world network is compact. 

\subsection{`Small worlds' is not the same as `small-world networks'}\label{ss-is_not}

The small-world networks are very specific graphs. 
In contrast, the small worlds, that is networks with a small-world effect, is an incredibly wide class of networks---practically all networks which we discuss. These networks are more compact than any finite-dimension lattice.

\subsection{Fat-tailed degree distributions}\label{ss-fat_tailed}

Actually, properties of classical random graphs do not differ tremendously from those of infinite-dimension lattices. However, if a network has a degree distribution with sufficiently slowly decreasing degree distribution, as in most important real-world networks, the difference is striking. 

Usually researchers try to fit empirical degree distributions by specific 
power-law dependences $P(k) \propto k^{-\gamma}$ (scale-free degree distributions). 
However, a far more important (and reliable) observed fact is that the higher moments of the empirical degree distributions diverge in large networks. 
This observation shows that, with noticeable probability, vertices of high degree are present in real networks, unlike classical random graphs. It is this presence that produces strong effects.

\subsection{Reasons for the fat-tailed degree distributions}\label{ss-reasons}

The main explanations of the fat-tailed form of empirical degree distribution are as follows:  
 
\begin{list}{}{\leftmargin=18pt}

\item[(1)]
{\em Self-organization} (or, rather, {\em self-organized criticality}): 
while evolving, a network self-organizes in a structure with an essential role of hubs.  

\item[(2)]
{\em Optimization processes involving many agents}: vertices arrange their connections in the optimal way. Actually this means an extensive competition of trade-offs, where each vertex `tries to find' the optimal combinations of numerous (often mutually contradictory) factors. In other words, everybody tries to arrange his or her connections in the best way taking into account different factors---everybody tries to make the best choice.  

\item[(3)]
{\em Multiplicative stochastic processes} may produce slowly decreasing distributions. (In multiplicative stochastic processes, variables change by random factors with time, that is the changes are relative.) 

\item[(4)] 
{\em The interconnection of geographically close vertices} may result in network architectures with fat-tailed degree distributions, Refs.~\cite{a02,rcbh02}. 

\item[(5)]
Fat tails of degree distributions may emerge as {\em a secondary effect}. 
Suppose that connections in a net are determined by a set of some intrinsic properties of vertices (`hidden variables'). 
The statistics of these hidden variables is explained by `external' reasons. 
Specific forms of the distributions of hidden variables result in slowly decreasing distributions of a vertex degree, see Refs.~[14--18]. 

\end{list}

Note that this classification is rather conventional. Rigid boundaries between explanations (1), (2), and (3) are absent.

\subsection{Preferential linking}\label{ss-preferential}

Maybe, the most popular self-organization mechanism is preferential attachment 
(preferential linking): vertices of high degree attract new connections with higher probability. 
In more precise terms, he probability that a new edge become attached to a vertex with $k$ connections is 
proportional to some function of $k$, a preference function, $f(k)$.\footnote{
In `inhomogeneous networks', the form of the preference function varies from vertex to vertex. 
}. 
The resulting structure of an evolving net is determined by the form of this function. 

Scale-free degree distributions may emerge only if the function $f(k)$ is linear, that is the probability of attachment is $(k+A)/(\langle k \rangle + A)$, where $A$ is some constant. It seems, this is a widespread situation in real networks. 
This form of preference usually produces the $\gamma$ exponents between $2$ and infinity. 

Models of evolving systems based on this concept was proposed by G.U.~Yule (1925) and H.E.~Simon (1955). To growing networks, this idea was applied by D.J.~de S.~Price (1976)---a linear preference function, and by A.-L.~Barab\'asi and R.~Albert (1999)---a proportional preference function. 

In particular, in {\em the Barab\'asi--Albert model}, a growing network is a so called `citation graph'. This means that new connections emerge only between a new vertex and existing ones. At each time step, a new vertex is added to the net and become attached to one or several vertices, which are chosen with proportional preference.

\subsection{Condensation of edges}\label{ss-condensation}

The preferential linking mechanism effectively works only in non-equilibrium networks. 
In more strict terms, only in non-equilibrium (e.g., growing) networks, linear preferential linking necessarily leads to scale-free architectures at any mean degree values. In equilibrium networks, even linear preference produces fat-tailed distributions only above some critical value of the mean degree. 

Below this point, that is in a more sparse network, the degree distribution is a rapidly decreasing function. Above this point, the condensation of edges takes place. In other words, a finite fraction of edges turns out to be attached to a vanishingly small fraction of vertices, or even to a single vertex. The degree distribution for the rest vertices is fat-tailed. 

Strong inhomogeneity of a network also can lead to the condensation of edges. 


\subsection{Cut-offs of degree distributions}\label{ss-cut-offs}

Clearly, in a finite size network, vertices of an infinitely large degree are absent. This means the presence of a size-dependent cut-off in the degree distribution, so that `perfect' scale-free degree distributions are impossible. 
In small networks, a cut-off obstructs the observation of fat-tailed distributions.  

The position of the cut-off depends on details of a network. 
E.g., in scale-free citation graphs of size $N$, the cut-off is 
$k_{\mbox{\footnotesize cut}} \sim N^{1/(\gamma-1)}$, 
in other situations, 
$k_{\mbox{\footnotesize cut}}$ 
may be of the order of $N^{1/2}$, etc.

\subsection{Reasons for correlations in networks}\label{ss-reasons_for_correlations} 

The main reasons for correlations in networks are as follows:  

\begin{list}{}{\leftmargin=18pt}

\item[(1)] 
Degree--degree correlations are the immediate result of the evolution of a network. In evolving network, there are only few exceptions, where degrees of the nearest-neighbour vertices are weakly correlated. (One of this exceptions is the Barab\'asi--Albert model.) 

\item[(2)]
Maximally random network is inevitably uncorrelated if multiple connections and loops of length one are allowed. However, if multiple connections and one-loops are forbidden, a maximally random network with a fat-tailed degree distribution may show strong correlations between the degrees of the nearest neighbours \cite{ms02}. 

\item[(3)]
Projections of uncorrelated multi-partied graphs are correlated networks. 

\end{list}

Let us explain the third possibility in more detail. Collaborations may be naturally described by so-called {\em bipartite graphs}, where the vertices of the first kind show collaborators, the vertices of the other type show the acts of collaboration, and edges connect each act of collaboration to all collaborators participating in this act. 

The same collaboration can be depicted by using only one type of vertices, namely collaborators. In this representation, two vertices are connected if they participate in at least one act of collaboration. The resulting graph is a one-mode projection of the bipartite collaboration network. 

The basic formal construction of a bipartite random network is {\em a generalized configuration model}, which is a direct generalization of the one-partite configuration model. In simple terms, this is a maximally random bipartite graph with two given degree distributions for both types of vertices. This is an uncorrelated bipartite network. However, its one-mode projection is a correlated network, which may have strong clustering, and may have degree--degree correlations. 

The last circumstance explains, in particular, why do real (one-mode) collaboration networks have strong clustering.

\subsection{Classical random graphs cannot be used for comparison with real networks}\label{ss-cannot_be_used}

We have again mentioned the configuration model, which is the simplest, basic complex random network. This model is well studied. If a degree distribution is fat tailed, the configuration model shows effects, crucially different from those of classical random graphs. 
Real networks are usually so different from classical random graphs that 
any comparison of real-world, complex networks with classical random graphs is meaningless. 
Instead, as a starting point, one must compare empirical data on a real network with the simplest complex uncorrelated network with the same degree distribution. This is the configuration model.

\subsection{How to measure degree--degree correlations}\label{ss-how_to_measure}

When an empirical researcher studies the statistics of vertex degree in a network, that is measures a degree distribution in the range $0 \leq k_{\mbox{\footnotesize max}}$, he or she has data for these $k_{\mbox{\footnotesize max}}$ points, obtained from inspection of $N$ vertices. When an empirical researcher studies the statistics of the correlations between the nearest-neighbours in a sparse network---the joint degree--degree distribution $P(k',k'')$, he or she has data for a much larger number $k_{\mbox{\footnotesize max}}^2$ points, obtained from only $\sim N$ edges. 
So, the statistics of the empirical joint degree--degree distribution are inevitably poor, and fluctuations will be strong on the resulting plot. Instead, empirical researchers usually describe these fluctuations by using a more coarse, but less fluctuating, characteristic. 
This is the mean degree $\overline{k}_{\mbox{\footnotesize nn}}(k)$ of the nearest neighbours of a vertex of degree $k$. This characteristics is can be easily expressed in terms of the joint degree--degree distribution.\footnote{ 
We, however, recommend that hindering fluctuations be reduced by using cumulative distributions: $P_{\mbox{\tiny cum}}(k',k'') = \sum_{q' \geq k',q'' \geq k''}P(q',q'')$.  
}

\subsection{Assortative and disassortative mixing}\label{ss-assortative}

Situations where vertices of high degrees mostly have the nearest neighbours of high degrees, i.e. where $\overline{k}_{\mbox{\footnotesize nn}}(k)$ grows with $k$, are called {\em assortative mixing} (this term is taken from sociology). 

The opposite case, where where vertices of high degrees mostly have the nearest neighbours of low degrees, i.e. where $\overline{k}_{\mbox{\footnotesize nn}}(k)$ decreases with $k$, is called {\em disassortative mixing}. 

For example, the Web and the Internet graph on the Autonomous Systems level show disassortative mixing. The Internet on the router level has week degree--degree correlations. Collaboration networks usually show assortative mixing.

\subsection{Disassortative mixing does not mean that vertices of high degrees rarely connect to each other}\label{ss-does_not_mean}

Rather counterintuitively, in a network where highly connected vertices mostly have neighbours with few connections, vertices of high degrees may turn out to be interconnected with high probability. 
  
This claim is illustrated by the following estimate. 
In a correlated unipartite network, the average number of edges between vertices of degrees $k'$ and $k''$ is $N(k',k'')=\langle k \rangle N P(k',k'')$. The maximum possible number of connections between these vertices of degrees $k'$ and $k''$ is $N(k')N(k'') = NP(k')NP(k'')$ (multiple connections are ignored). So, the fraction 
\\
$N(k',k'')/[N(k')N(k'')] = [\langle k \rangle P(k',k'')]/[NP(k')P(k'')]$ of the possible connections is present. 
In simple terms, this is the probability that an edge between a pair of vertices of degrees $k'$ and $k''$ is present. 
For instance, in an uncorrelated network, 
the resulting number is $k' k''/(N\langle k \rangle)$ and may be large enough at large degrees. This is the simplest case, but evidently, similar conclusions are valid for both the assortative and disassortative types of correlations. 

For example, in the Internet, the set of Autonomous Systems with highest numbers of connections is practically fully interconnected.

\subsection{Reciprocal links in directed networks}\label{ss-reciprocal}

For simplicity, in our discussions we neglect one detail---multiple 
connections. However, sometimes, the presence of multiple connections may be important. 
Often, multiple connections in networks are considered as something exotic. 
The counterexample is the WWW---an directed network, where about $30$ per cent of hyperlinks have opposite-directed ones. That is, if one page has a reference to another page, the latter refers to the former with high probability. The same occurs in directed email networks.


\subsection{Ultra-small-world effect}\label{ss-ultra_small}

In an uncorrelated network with an arbitrary degree distribution $P(k)$, the degree distribution of the nearest neighbour of a vertex is $kP(k)/\langle k \rangle$, which is quite different from $P(k)$.\footnote{
A similar effect takes place in uncorrelated networks. 
} 
This principal difference is the origin of many effects in complex networks. 

The mean degree of the nearest neighbour of a vertex is $\langle k^2 \rangle/\langle k \rangle$, which is greater than the mean degree of the vertex in the network. This circumstance changes the formula (\ref{1}) for the mean shortest path length to the following one:

\begin{equation}
\overline{\ell} \approx \frac{\ln N }{\ln [(\langle k^2 \rangle/\langle k \rangle) - 1]}
\, .  
\label{8}
\end{equation} 
One can see, that if the second moment of the degree distribution diverges in the infinite network limit, the average number of the second nearest neighbours of a vertex approaches infinity, and the formula (\ref{8}) cannot be used. 
In this situation, $\overline{\ell}(N)$ grows with $N$ slower than $\ln N$, which may be called `the ultra-small-world' effect.

\subsection{The importance of the tree ansatz}\label{ss-tree_ansatz}

Formulae (\ref{1}) and (\ref{8}), and many other basic results were obtained by using {\em the tree ansatz}. This is an assumption that a sufficiently vast environment of each vertex of a network has a tree-like structure. In more precise terms, it is assumed that as the total number of vertices tends to infinity, any finite environment of a vertex almost surely does not contain loops. In this case, loops occur only if the remote environment of a vertex is added. 
In particular, the tree ansatz is valid for large uncorrelated networks. 

For many characteristics of a network, remote environments of vertices are not important, and the simplifying tree ansatz may be used. Moreover, some of results obtained in the frames of the tree ansatz are still valid for networks with numerous loops and strong clustering.

\subsection{Ultraresilience against random failures}\label{ss-ultraresilience}

If the second moment of the degree distribution diverges (this, e.g., happens in infinite scale-free networks with $\gamma \leq 3$), the average degree of the nearest neighbour of a vertex also diverges. This means that a vertex in the {\em infinite network}, in average, has an infinite number of the second nearest neighbours. Evidently, this indicates the presence of the giant connected component in the network. 

Let us remove, at random, a finite fraction of vertices or edges from the network (random failure). One can see that after this removal, the average number of the second nearest neighbours of a vertex is still infinity. In other words, the random removal of any finite fraction of vertices or edges does not eliminate the giant connected component. In this situation, even the random removal of $99,99$ per cent of vertices or edges does not destroy the `core' of the network. That is, these networks are ultraresilient against random failures.

\subsection{When correlated nets are ultraresilient}\label{ss-when_ultraresilient}

The above claims have been made for uncorrelated networks. Nevertheless, one can show that in networks with correlations between degrees of the nearest neighbours, the average number of the second nearest neighbours of a vertex diverges in the same situation, that is when the second moment of the degree distribution diverges. 

So, the condition for ultraresilience against random failures is the same both for the correlated and uncorrelated networks: the second moment of the degree distribution must diverge.

\subsection{Vulnerability of complex networks}\label{ss-vulnerability}

On the other hand, it is highly connected vertices that enable the existence of the giant connected component in networks with fat-tailed degree distributions. This is only a small fraction of the total number of vertices. 
So that, an intentional damage of a network may have a strong effect, if one removes vertices of the highest degrees.   
In this case, it is sufficient to remove a small fraction of vertices to eliminate the giant connected component.

\subsection{The absence of an epidemic threshold}\label{ss-epidemic_threshold}

Another side of the ultraresilience against random failures is the absence of the epidemic threshold at the same condition. That is, in simple terms, any finite (nonzero) rate of the infection of the nearest neighbours of a vertex in a network with diverging second moment of the degree distribution leads to a `global epidemic'. 

Actually, the problems of the spread of diseases and random failures (or the percolation problem) are closely related. A low infection rate in the spread of diseases corresponds to a high fraction of removed vertices or edges in the percolation problem. The pandemic corresponds to the presence of the giant connected component in the randomly failed network. 
So, the absence of the epidemic threshold corresponds to the impossibility to eliminate the giant connected component of the network by random removal of vertices or edges.

\subsection{Search based on local information}\label{ss-search}

Another effect of the divergence of higher moments of degree distribution is a diminishing of the characteristic time of a so called `local search' 
\cite{alph01}.  
Suppose that each vertex contains a full information about its nearest neighbours. In the local search problem, one must find a vertex with some desired information. 

A possible search strategy looks as follows. Start from an arbitrary vertex, and move along edges randomly, from vertex to vertex.   
Then, recalling that the mean degree of the nearest neighbour of a vertex is $\langle k^2\rangle/\langle k\rangle$, the typical search time is estimated as 
$N/(\langle k^2\rangle/\langle k\rangle)$. Consequently, this search is quick in networks with large second moment of the degree distribution.

\subsection{Ultraresilience disappears in 
finite networks}\label{ss-finite_size}

The ultraresilience against random failures and the absence of the epidemic threshold are determined by the divergence of the second moment of a degree distribution. This divergence is possible only in infinite networks. If a net is finite, the degree distribution necessarily has a size-dependent cut-off, and all the moments of the degree distribution are finite. In this case, the average number of the second-nearest neighbours of a vertex is finite, the giant connected component can be removed by random removal of a sufficiently large, but finite, fraction of vertices, and the epidemic threshold exists. 

Thus, for the observation of the ultraresilience and of the disappearance of the epidemic threshold, a network must be very large. 
Recall that as a rule, real networks are small.

\subsection{Critical behavior of cooperative models on networks}\label{ss-critical_behavior}

We have explained that networks are infinite dimension objects. Physicists know that critical fluctuations in cooperative models on infinite dimension objects are absent, and so the critical behavior is described by mean-field theories. 

So, the critical phenomena in networks should be described by mean-field theories. Indeed, in the case of equilibrium networks with degree distributions with a well-defined scale, the standard mean-field theory, with standard critical exponents, is valid. 

On the other hand, if the degree distribution of an equilibrium network is fat-tailed, critical behavior is non-standard. This implies unusual values of critical exponents. Moreover, the order of a phase transition may be high and even approach infinity. Nonetheless, the critical fluctuations are absent even in this case, and a mean field theory still works. The point is that the mean-field behavior is non-standard because of the presence of highly connected vertices in the network.

The above general claims are valid for various cooperative phenomena (percolation, magnetic phase transitions, etc.) in various networks (correlated and uncorrelated networks, small-world networks, etc.)\footnote{
Note, however, a principal difference from the synchronization phenomenon: synchronization is possible even in a system of two coupled oscillators, and `normal' phase transitions are realised in infinite cooperative systems.  
}.

\subsection{Berezinskii-Kosterlitz-Thouless phase transitions in networks}\label{ss-bkt}

There is an impressive exception from the above scenario.  
In some situations, nonequilibrium networks may have a `non-mean-field' phase transition with {\em the Berezinskii-Kosterlitz-Thouless singularity}. Near this infinite-order phase transition, the order parameter (e.g., the size of the giant component in percolation-like problems) changes as $\exp(-\mbox{const}/\sqrt{b-b_{\mbox{\footnotesize c}}})$. Here $b$ is some control parameter, and $b_{\mbox{\footnotesize c}}$ its critical value.

\subsection{Cascading failures 
}\label{ss-cascading}

We have explained that self-organized-criticality mechanisms can produce complex network architectures. 
On the other hand, one can study avalanches and other phenomena associated with self-organized criticality on networks, e.g., a sand-pile problem on networks, Ref. \cite{lgkk04}. The impressive effect of cascading failures in power grids explains the wide interest in avalanche phenomena on networks.  

In the simplest model \cite{w02}, cascades on networks are related not to self-organized criticality but rather to percolation problems and to the spread of infections. The model illustrates the phenomenon of global cascades induced by a local perturbation on networks. In this model, a vertex may be in two states: $A$, which is the initial state and $B$, which is the perturbed state. The dynamics of the model is described by the following rule: a vertex adopts state $B$ if at least a fraction $p$ of its nearest neighbours are in state $B$. Here the threshold $0<p<1$ is the parameter of the problem. Otherwise, vertex remains in state $A$. This process indeed resembles the spread of infections, and the global cascades correspond to pandemics.   
It turns out   
that only in some restricted range of $p$ and network structure parameters, global cascades are possible.

\subsection{Cliques and communities}\label{ss-cliques_and_communities}

Cliques are fully connected subgraphs of a graph. 
Communities are (rather poorly defined) subgraphs, where vertices are `better' connected to each other than the full set of the vertices of a network. For practical purposes, it is often important to find the full set of communities in the network. In principle, this problem has no unique solution. One of the numerous approaches for indexing the communities is based on inspecting the structure of the shortest paths in the network (M.~Girvan and M.E.J.~Newman).

\subsection{Betweenness}\label{ss-betweenness}

How are shortest paths between the pairs of vertices distributed over the network? This distribution is characterized by {\em betweenness}. Let the total number of the shortest paths between vertices $i$ and $j$ be $B(i,j)$ and $B(i,m,j)$ of them pass through vertex $m$. The betweenness $b(m)$ of the vertex $m$ is 

\begin{equation}
b(m) = \sum_{i \neq j}\frac{B(i,m,j)}{B(i,j)}
\, .  
\label{9}
\end{equation} 
In simple terms, this is the probability that a shortest path between a pair of vertices of a network passes through the vertex $m$. 
In a similar way one can define an edge betweenness. Unlike degree, the value of the betweenness of a vertex reflects the topology of the entire graph.   
Evidently, vertices (edges) with high betweenness (a high fraction of passing shortest paths) play especially important role in a network. 

\subsection{Extracting communities}\label{ss-extracting}

In the Girvan-Newman algorithm (see, e.g., Ref. \cite{ng03}), edges with maximal betweenness in the network are deleted one by one. This deletion changes the structure of the shortest paths in the network, and so a betweenness of each edge is recalculated after each deletion. At some step, the network turns out to be divided into two clusters---two largest communities, and so on. The result is a tree, where smaller communities are included in larger ones. 
The distribution of the sizes of resulting communities in many real networks is a power law.


\subsection{Optimal paths}\label{ss-optimal_paths}

A network shows the  
small-world effect if  
the mean shortest-path length $\overline{\ell}$ grows slower than any power of the number of vertices, $N$. Let us remove a fraction of vertices or edges from a network with the small-world effect, so that we approach the percolation threshold from above. In other words, let us nearly eliminate the giant connected component of the network. The mean shortest-path length is determined by the shortest paths between vertices in the giant connected component (in total, $N_g$ vertices). It turns out that near the point of the birth of the giant component, the $\overline{\ell}(N_g)$ dependence is a power law. 

Instead of removing edges, one may ascribe them random weights and consider the optimal paths between vertices. Here the optimal path is the shortest one taking into account the weight of edges. If the disorder (variations of the weight) is strong, the size dependence of the optimal path become power-law \cite{bbchs03}. This is a way to eliminate the small-world effect.

\subsection{Distributions of the shortest-path length and of the loop's length are narrow}\label{ss-narrow}

On can show that, with few exceptions, large networks with the small-world effect have a very narrow distribution of of the shortest-path length. As the size of a network grows to infinity, the ratio of the width of the distribution of the shortest-path length and the mean shortest-path length approaches zero. That is, in the thermodynamic limit, this distribution is the delta-function. 

The same is valid for the distribution of the loop's length in large networks with the small-world effect \cite{rkbb04}.

\subsection{Diffusion on networks}\label{ss-diffusion}

Thus, vertices of large networks are almost surely mutually equidistant. 
This claim has a number of immediate consequences. For example, consider a diffusion process on a large network. 
A particle, which initially was at vertex $0$, jumps from a vertex to a vertex. 
At what distance from vertex $0$ will the particle be at infinite time? 
The mutual equidistance of vertices guarantees that with the great probability, the particle will be at the distance $\overline{\ell}$ from 
the starting vertex. 

Note that 
two different diffusion problems can be considered on networks, but for both of them the above claim is valid.  
In the first problem, the probability that the particle leaves a vertex per time step is fixed. Then at infinite time, the probability that a particle will occur at a vertex, is proportional to the degree of this vertex. 
In the second diffusion problem, the probability that the particle moves to a given nearest neighbour per time step is fixed. 
Then finally, the particle will occur at each vertex of the net with equal probability.


\subsection{What is modularity?}\label{ss-modularity}

The `modularity' and the 'modular structure' of networks are frequently used terms, but what do they precisely mean? 
Assume that a network may be divided into modules labeled by an index $i$, so that the full set of modules is $\{i\}$.
Let $e_{ij}$ be the fraction of edges in the network that connects modules $i$ and $j$. Then, $e_{ii}$ is the fraction of edges in the network that are inside of module $i$, and $n_i = \sum_j e_{ij}$ is the fraction of edges that are attached to the vertices of module $j$. 
Note that $\sum_j$ includes the term with $j=i$. 
The modularity \cite{ng03} for this specific division of a network into modules is 

\begin{equation}
M[\{i\}] = \sum_i e_{ii} - \sum_i n_i^2 
\, .  
\label{10}
\end{equation} 
One can check that $M$ can take values between $0$ and $1$. If edges connect vertices irrespectively to this division into modules, then $M=0$. With increasing  $M$, the division into modules becomes more pronounced. 

Note that the modularity (\ref{10}) is defined only for a given set of 
modules of a network. Another division of the network produces a different value of the modularity. 
 In principle, one can define the modularity of a network as the maximum value of $M$ for all possible sets of modules. Unfortunately, the computation of this maximum is a really hard problem.

\subsection{Hierarchical organization of networks}\label{ss-hierarchical} 

This is another popular and yet poorly defined term. 
One of possible ways to characterize the hierarchical structure of a network is based on the notion of a {\em hierarchical path} \cite{g01}. 

A path between nodes $a$ and $b$ is hierarchical if (1) the degrees of verti\-ces along this path 
vary monotonously from one vertex to the other or (2) vertex degrees first monotonously grow, reach maximum value, and then monotonously decrease along the path. 
Let the fraction $H$ of the shortest paths in a network be hierarchical. 
Then this number $H$ can be used as a metric of a hierarchical topology \cite{tmms03}. If $H$ of a network is sufficiently close to $1$, the network has pronounced hierarchical organization. For example, the Internet at the Autonomous Systems level has $H \approx 0.95$.  

In a number of papers, 
a specific degree dependence of the local clustering $C(k)$ was treated as a direct indicator of 
the hierarchical organization of a network. We believe that in general, this association is incorrect. At least, the form of $C(k)$ does not correlate with the value of $H$.

\subsection{Convincing modelling of real-world networks: Is it possible?}\label{ss-convincing}

We have indicated several different ways to generate networks with complex architectures. But the aim is to explain how the complex architectures of real networks emerge. 
In principle, by using a sufficiently large number of fitting parameters of, e.g., self-organizing models, one can satisfactory describe a set of empirical characteristics or real networks. However, this `successful' description 
has no explanatory power. 

A really convincing model of a real network must explain a sufficiently large set of empirical data without fitting. The model parameters must be expressed in terms of only known basic numbers of a network (e.g., the total numbers of vertices and edges, etc.), some input rates (e.g., the rate of the network grows), or be expressed in terms of known `external' factors if they influence the evolution of networks. So that, to be convincing one has to avoid fitting. 
The question is: is it possible to describe so complex systems without any fitting?

\subsection{The `small' Web}\label{ss-small_web}

We stressed that even large real-world networks are mesoscopic objects. 
Even the extremely large Web contains only about $10^{10}$ (sufficiently `static') pages. Moreover, the total volume of information on the Web is not so great---only about 200 Terabytes on the `surface', `static' Web. It is this Web that is explored by search engines. 
Due to the smallness of the surface Web, Google can store a large number (many dozens) of the Web copies on the hard drives of Google servers.

\subsection{The failures and perspectives of the physics approach to complex networks}\label{ss-failures}


Networks are widespread objects with properties remarkably different from those of lattices.   
Physicists in the science of networks use traditional, effective methods of statistical mechanics and are involved in the empirical research of real-world networks.  
This 
has allowed them to find new classes of networks, very common in the real world, and understand a number of their features. 

The problem is that this new understanding of complex networks by physicists still did not produce significant practical results. We believe that a more practical, 
applied direction provides an encouraging perspective for the network research.

\subsection{A remark about references}

The list of references mostly contains large reviews and reference books. The readers can find detailed bibliography in these sources. In addition, several quite recent papers are included in the list. 
For more detailed and systematic introduction to the topic of complex networks, the readers may refer to our book \cite{dmbook03}.

\subsection*{Acknowledgements}

The authors thanks M.~Alava, A.V.~Goltsev, J.G.~Oliveira, A.M.~Povolotsky, and A.N.~Samukhin for useful discussions. The authors were partially supported by project POCTI/MAT/46176/2002.







\label{s-references}


\begin{small}

\end{small} 


\end{document}